\documentclass[aip,pof]{revtex4-1}
\usepackage{amsmath}
\usepackage{multirow}
\usepackage{graphicx,subfigure}
\usepackage{xfrac}
\begin{document}


\title{Optimization of micropillar sequences for fluid flow sculpting} 



\author{Daniel Stoecklein}
\affiliation{Department of Mechanical Engineering, Iowa State University}

\author{Chueh-Yu Wu}
\affiliation{Department of Bioengineering, University of California at Los Angeles}

\author{Donghyuk Kim}
\affiliation{Department of Bioengineering, University of California at Los Angeles}

\author{Dino Di Carlo}
\affiliation{Department of Bioengineering, University of California at Los Angeles}

\author{Baskar Ganapathysubramanian}
\affiliation{Department of Mechanical Engineering, Iowa State University}


\date{\today}

\begin{abstract}
Inertial fluid flow deformation around pillars in a microchannel is a new method for controlling fluid flow.  Sequences of pillars have been shown to produce a rich phase space with a wide variety of flow transformations. Previous work has successfully demonstrated manual design of pillar sequences to achieve desired transformations of the flow cross-section, with experimental validation. However, such a method is not ideal for seeking out complex sculpted shapes as the search space quickly becomes too large for efficient manual discovery.  We explore fast, automated optimization methods to solve this problem.  We formulate the inertial flow physics in microchannels with different micropillar configurations as a set of state transition matrix operations. These state transition matrices are constructed from experimentally validated streamtraces for a fixed channel length per pillar. This facilitates modeling the effect of a sequence of micropillars as nested matrix-matrix products, which have very efficient numerical implementations. With this new forward model, arbitrary micropillar sequences can be rapidly simulated with various inlet configurations, allowing optimization routines quick access to a large search space. We integrate this framework with the genetic algorithm and showcase its applicability by designing micropillar sequences for various useful transformations. We computationally discover micropillar sequences for complex transformations that are substantially shorter than manually designed sequences. We also determine sequences for novel transformations that were difficult to manually design. Finally, we experimentally validate these computational designs by fabricating devices and comparing predictions with the results from confocal microscopy.\end{abstract}


\maketitle 


\section{Introduction}

The physics of inertial fluid flow deformation at the microscale has seen a surge of theoretical and experimental interest in the past decade\cite{Amini2014}.  Use of the inertial flow regime ($1 < Re < 100$, with the Reynolds number $Re = \frac{\rho V D_H}{\mu}$, where $\rho$, $V$, and $\mu$ are the fluid density, average downstream velocity, and viscosity, and $D_H$ the hydraulic diameter) in the microfluidics community contrasts the previously held notion that most practically useful flows were in the Stokes regime\cite{Stone2004} ($Re \rightarrow 0$), though the effects of inertial fluid flow has been observed since the 1960s\cite{Segre1961}.  New applications have since emerged, with inertial focusing in particular, leading to new methods for high-throughput cytometry\cite{Bhagat2008,Oakey2010,DiCarlo2010_1,Goda2012,DiCarlo2011,Hansson2012,Ciftlik2013}.

\begin{figure}[t]
\includegraphics[width=\textwidth, keepaspectratio]{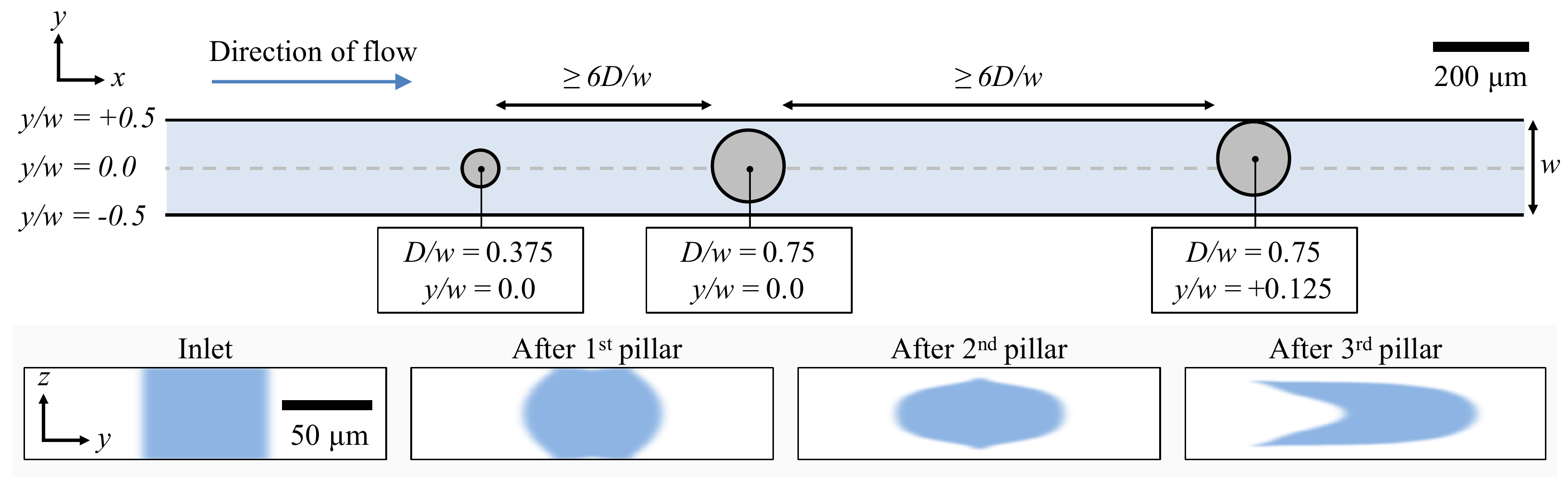}
\caption{Schematic of a micropillar sequence (drawn to scale).  Note the varying inter-pillar spacing based on the size of the preceding pillar, allowing the turning motion from each pillar's deformation to saturate before the flow arrives at the next pillar.  All lengths are normalized to the microchannel width, $w$.  Below are cross-sectional images of sculpted fluid based on the sheathed flow shown in the inlet image and the micropillar sequence as illustrated.  The predictions come from the simulation method described in this work ($Re=20$).\label{fig:schematic}}
\end{figure}

More recently, fluid sculpting via inertial flow has been demonstrated through so-called ``pillar programming''\cite{Amini2013,Stoecklein2014}.  See Fig~\ref{fig:schematic} for a general schematic of pillar programming.  In pillar programming, pillars spanning the height of a microchannel create a secondary flow which can be used in sequence with additional downstream pillars to generate a net deformation to the fluid.  The idea of programmability comes from the micropillars being spaced far enough apart to allow for their individual deformations to saturate.  Thus, the flow sculpted by one pillar can be used as an input to another pillar, without concern for cross-talk or time dependent effects.  This allows for a powerful computational shortcut for future simulations, as the deformation for particular micropillar configurations need be computed only once, with the resulting fluid displacement forming a 2D ``advection map''.  Net deformation for a sequence of micropillars can then be rapidly simulated using the pre-computed advection maps for each micropillar in the sequence.  This subverts the need to solve Navier-Stokes equations over large 3D domains for entire microchannel designs using pillar programming.

Amini, et al determined numerically that an inter-pillar spacing of 6 pillar diameters is sufficient to prevent upstream flow effects of one pillar from interacting with a previous pillar's deformation\cite{Amini2013}.  In experiments, they use a spacing of 10 pillar diameters to ensure the pillar deformations are completely isolated.  Therefore, each pillar in a designed microfluidic device will extend the length of the microchannel by a fixed amount, and its pillar programming simulation is valid only in the flow conditions for the pre-computed advection map.

\begin{figure}[h]
\includegraphics[width=\textwidth, keepaspectratio]{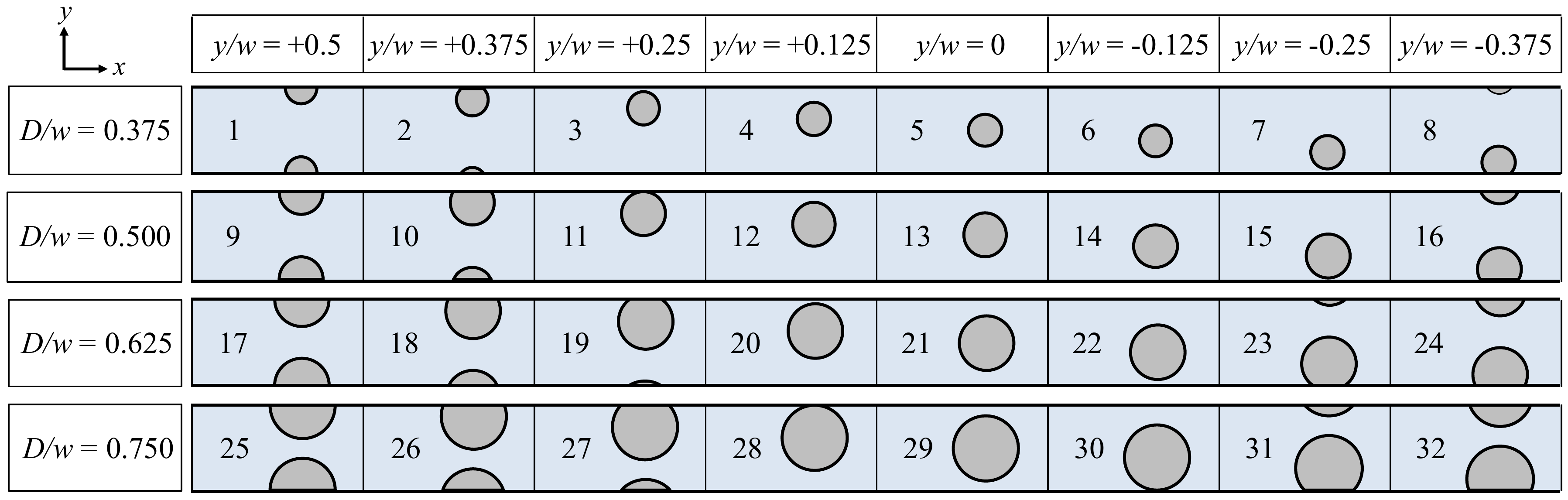}
\caption{Set of previously used micropillar configurations\cite{Amini2013, Stoecklein2014, Diaz-montes2014}.  Each pillar spans the height of the microchannel.  Note the periodic boundary condition, for which a pillar being placed so close to the microchannel wall will have the merged volume protrude from the opposite wall.  For this paper, each pillar configuration of diameter $D/w$ and offset $y/w$ has an integer index, which is shown next to each depiction of the micropillar.}  \label{fig:pillar_index_table}
\end{figure}

With a coarse but broad design space for initial exploration, a set of widely varied transformations was found and experimentally validated using this method for forward simulation\cite{Stoecklein2014}.  The designs used 1 to 10 micropillars, each chosen from the set of configurations shown in Fig~\ref{fig:pillar_index_table}.  A user-friendly GPU based software platform ``uFlow'' was developed in tandem with this work and made freely available, enabling any researcher with modest computing power to manually test various micropillar sequence designs for their own purposes.  uFlow is built on top of computationally expensive solutions to the Navier-Stokes equations\cite{Stoecklein2014}, but as a standalone product it uses lightweight, pre-computed advection maps, relieving the end-user of time consuming calculations.  Using this framework, manual pillar programming has been successfully employed for shaping polymer precursors for streams\cite{Nunes2014} and particles\cite{Paulsen2015,Wu2015}, reducing inertial flow focusing positions \cite {Chung2013}, and solution transfer around particles \cite{Sollier2015}.

Such manual exploration of flow deformations will quickly run up against the combinatorial phase space of possible pillar combinations\cite{Stoecklein2014}. For example, consider creating pillar sequences where each pillar is chosen from a set of 32 possible pillars of varying size and transverse location\cite{Diaz-montes2014} (see Fig~\ref{fig:pillar_index_table}). For a 10-pillar sequence, there are $32^{10} \approx 10^{15}$ possible pillar sequences, in addition to multiple inlet fluid flow configurations.  It becomes non-trivial (in terms of time and effort) to manually design a micropillar sequence by searching this phase space.  Even if manual design is attempted, a more efficient sequence with fewer pillars might be possible.  This is important in microfluidic devices, as the number of pillars in a sequence determines the length of the channel, and therefore the footprint of the device itself.  Longer channels and more pillars increase the pressure required for operation, which can introduce device flexure, and exacerbate transverse mass diffusion that increases with fluid distance traveled\cite{Ismagilov2000}.  Decreasing the channel length can not only reduce the pressure drop to improve the performance of the device, but also provide more space downstream for applications, for example, complex 3D shaped particle fabrication or particle separation and solution exchange\cite{Nunes2014,Paulsen2015,Sollier2015,Wu2015}.  Therefore, an optimization scheme is needed to enhance the capability of pillar programming while still remaining accessible to, and easily implementable by, an interested researcher with modest computing hardware. Although pillar diameter will play a role in the required pressure to drive flow through a micropillar sequence, we have determined that microchannel length has a greater overall impact on pressure drop (more information available in supplemental material\cite{Note3}).  Hence, the first priority in optimization is to find a flow shape that matches the target.  This is the motivation for the work presented here.  Secondary to this is the minimization of micropillar sequence length (determined primarily by the number of micropillars needed), which contributes advantages in terms of both reduced pressure drop and space for applications.  Tertiary goals, such as optimizing for the effects of pillar diameter on pressure drop, are currently not taken into consideration for optimization.  

We show two steps toward a framework for automated pillar sequence design: the development of an efficient forward model for rapid evaluation of arbitrary micropillar sequences, and the formulation of the design problem as an optimization problem.  To accomplish these steps, we first convert the per-pillar advection maps to sparse transition matrices, which store fluid state displacement information as transition probabilities across states.  These matrices can be multiplied together to quickly propagate net deformation across many pillars.  This reduction of complex three-dimensional deformation simulations to numerically efficient matrix-matrix multiplication subsequently enables effectively pairing with a genetic algorithm for optimal design\cite{Mott2009} of micropillar sequences.  To demonstrate the effectiveness of this platform, we show the design and experimental validation of optimized micropillar sequences for previously discovered deformations, as well as novel designs which have no previously known pillar sequences.

\begin{figure}[t!]
\includegraphics[width=\textwidth, keepaspectratio]{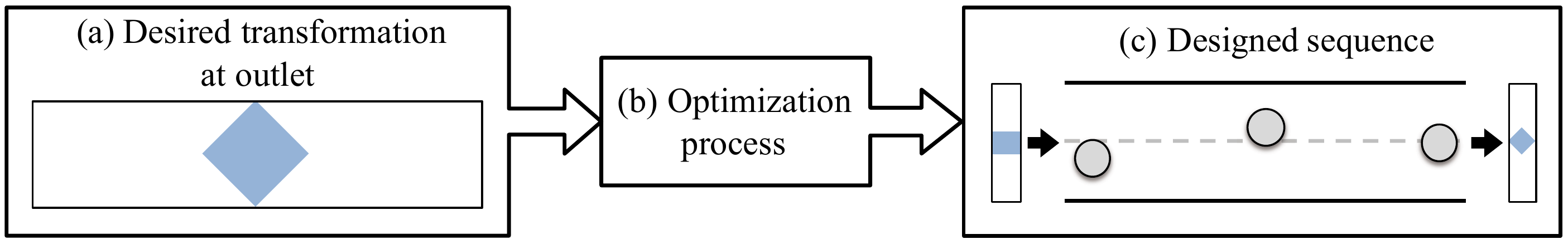}
\caption{Illustration of the design problem, where a micropillar sequence is desired that will deform fluid into the given fluid flow shape (a).  Some optimization routine (b) must determine a micropillar sequence (c) that yields a flow shape that closely matches the given fluid flow shape.  \label{problem_formulation}}
\end{figure}

\section{Problem formulation}

The goal is to develop a computational framework that accepts a desired transformation as an input, and produces a micropillar sequence design for the nearest possible match to this transformation as an output (see Fig.~\ref{problem_formulation}).  For this work, the simulation and fabrications parameters will remain within the constraints of pillar programming as previously defined.

\subsection{Forward model}

The design problem potentially requires evaluation of many thousands of different pillar sequences. We refer to each such evaluation as solving the forward model (in contrast to the design problem). A forward model must therefore be accurate and fast.  We begin by creating advection maps from streamtrace data for various pillar configurations, which is computed from our validated in-house finite element Computational Fluid Dynamics (CFD) framework\cite{Amini2013,Stoecklein2014,Diaz-montes2014}.  We then translate the displacement information from the maps into transition matrices, which are an effective method for fast and accurate pillar program simulation.

\subsubsection{CFD data to advection maps to transition matrices}

\begin{figure}[h]
\includegraphics[width=\textwidth, keepaspectratio]{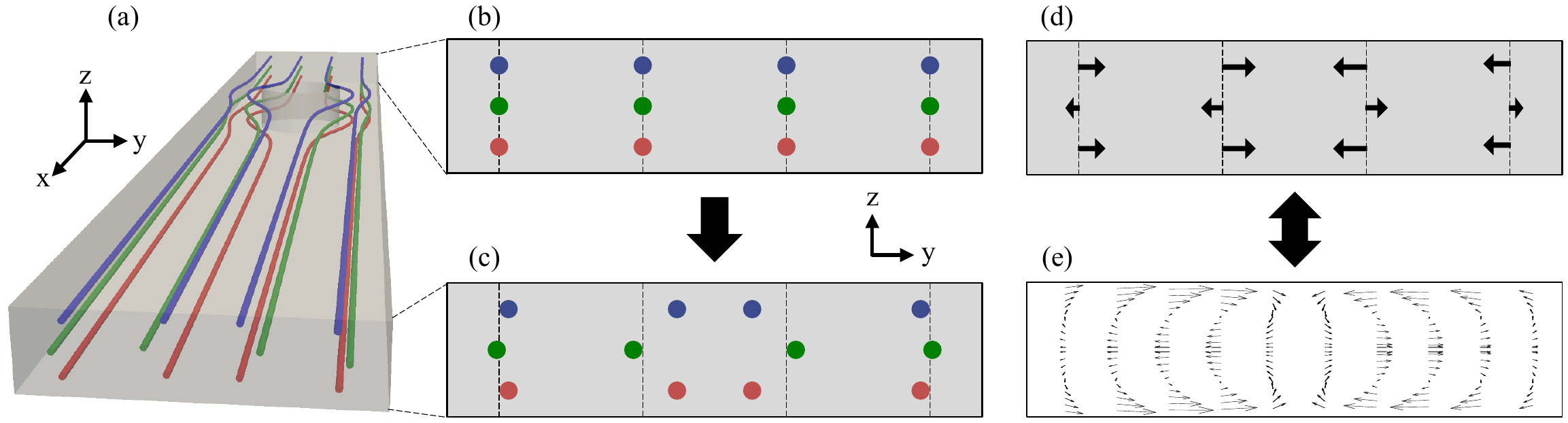}
\caption{(a) Streamtraces through a 3D velocity field are used to form advection maps by comparing inlet (b) and outlet (c) positions of infinitesimal massless, neutrally buoyant particles (in this case, $N_Y=4$, $N_Z=3$ particles are used).  (d) The resulting advection map shows the net secondary flow for a particular pillar configuration and flow condition, which has resolution determined by the number of particles used. (e) Shows a representative realistic 2D advection map. \label{streamtrace}}
\end{figure}

We start with a dataset of 3D velocity fields calculated for a set of individual micropillar configurations at $Re=20$ (See Fig~\ref{fig:pillar_index_table}).  The Navier-Stokes equations are solved (using the finite element method) in a domain that spans six diameters upstream and downstream of the pillar. A detailed description of this procedure is provided in our earlier work\cite{Stoecklein2014,Amini2013} ~and is not repeated here for the sake of brevity. Each 3D velocity field  is contracted to a 2D advection map by streamtracing uniformly distributed, neutrally buoyant particles through the velocity field (see Fig.~\ref{streamtrace}(a)).

\begin{figure}[t]
\includegraphics[width=\textwidth, keepaspectratio]{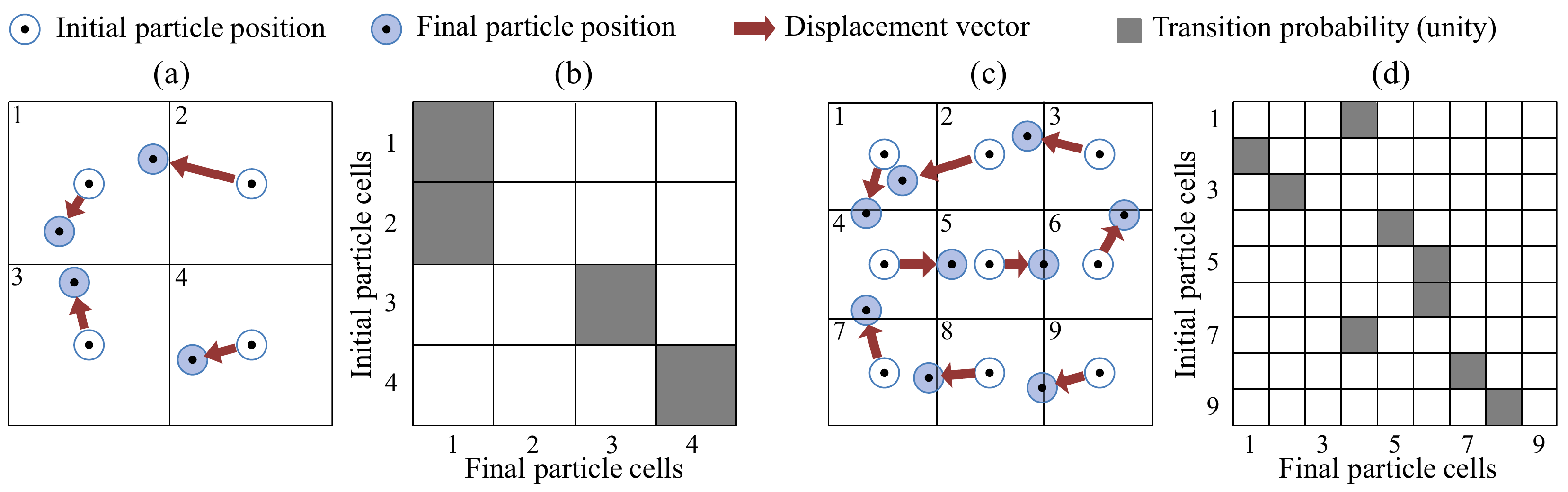}
\caption{(a) Mapping of a 4 particle advection map with $N_Y=2$, $N_Z=2$ to a 4$\times$4 transition matrix (b).  Each row corresponds to one cell. For each row, the column that is darkened represents the destination cell.  Increasing the number of particles in the advection map results in a more resolved transition map as shown in (c) and (d).\label{truncation_error}}
\end{figure}

For enhanced numerical efficiency, our idea is to make pillar programming amenable to using the BLAS (Basic Linear Algebra Subprograms) libraries within the CPU and/or GPGPU (General Purpose computation on Graphics Processing Units).  BLAS libraries are freely available software that are architecture-aware, utilize memory optimally, and are specifically pipelined and tuned for fast matrix-matrix operations.  These libraries are platform agnostic, and are usually efficiently implemented on most CPUs and GPUs.  Adapting pillar programming to such matrix-matrix operations allows for utilization of a growing repertoire of highly efficient optimization methods.

\begin{figure}
\includegraphics[width=\textwidth, keepaspectratio]{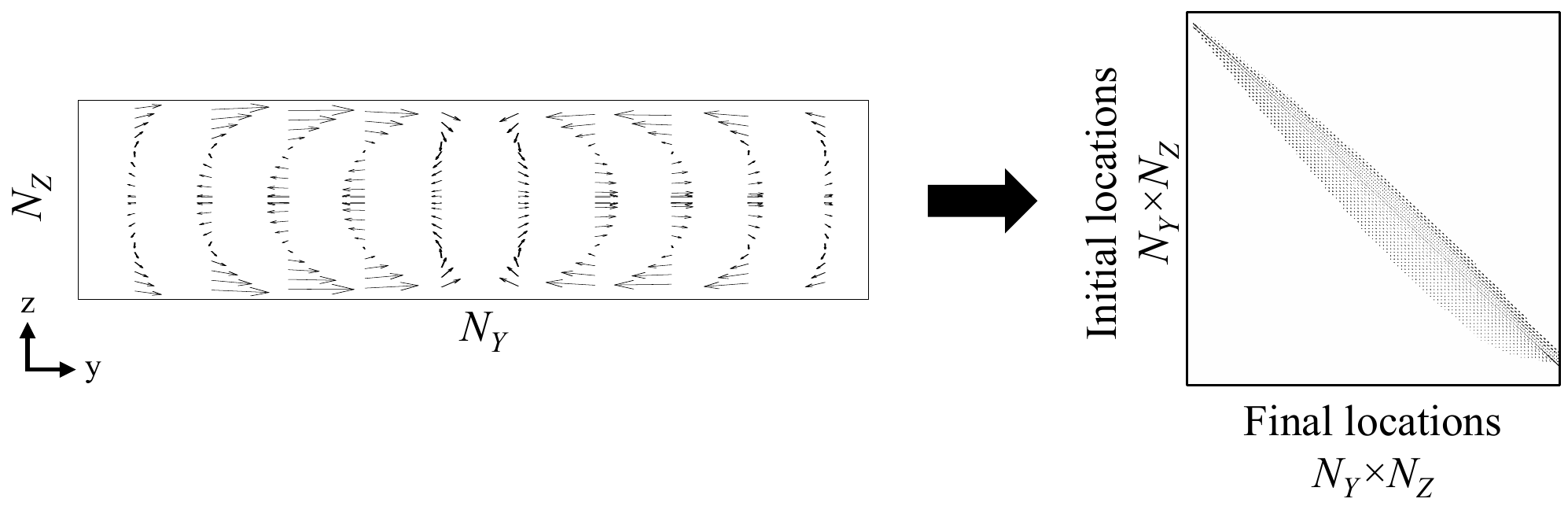}%
\caption{Advection maps converted to transition matrices. This matrix is sparse. The figure shows the non-zero entries of the matrix in black. \label{advection_demo}}%
\end{figure}

We convert the displacement information contained in an advection map into a sparse ``transition matrix''. This is accomplished by first discretizing the advection map into a finite set of $N$ cells, and subsequently identifying where each cell advects to (or is displaced to). Consider an advection map constructed by using  $N = N_Y \times N_Z$ particles that are uniformly distributed across the cross-section (with $N_Y$ particles along the $y$ direction, and $N_Z$ particles along the $z$ direction). We discretize the channel cross-section into $N= N_Y \times N_Z$ cells, with each cell center displacement given by the corresponding particle displacement from the advection map. This is illustrated in Fig.~\ref{truncation_error}(a) and Fig.~\ref{truncation_error}(c). 

The transition matrix is simply a matrix with $N$ rows. Each row of this matrix accounts for the transition information of one cell. Each row stores an individual cell's displaced location index (i.e. the index of the cell where this cell lands) as indicated by the advection data. Examples of such transition matrices are illustrated in Fig.~\ref{truncation_error}(b) and Fig.\ref{truncation_error}(d). This results in a sparse matrix that is easily stored and manipulated. Fig.~\ref{advection_demo} illustrates the sparsity of a representative transition matrix. 

\begin{figure}[t]
\includegraphics[width=\textwidth, keepaspectratio]{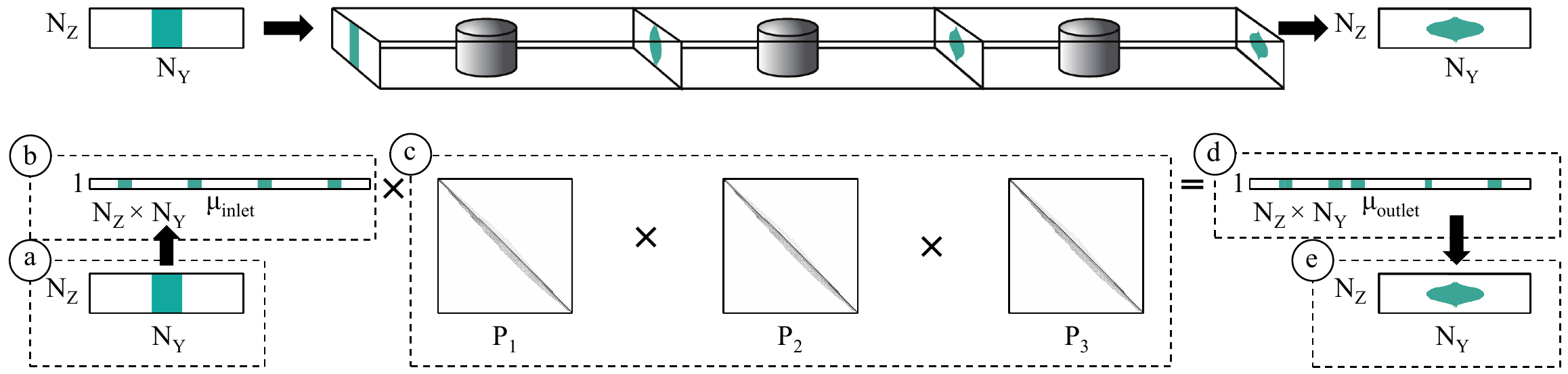}
\caption{Illustration of how transition matrices $P_1$, $P_2$, and $P_3$ can be used to simulate the net deformation from three individual micropillars.  First, the inlet flow condition (a) is reshaped to a row vector $\mu_{inlet}$ (b) with a length matching the dimension of the square transition matrix.  This vector is then multiplied by the product of the transition matrices (c), which forms an outlet row vector $\mu_{outlet}$ (d).  This can then be reshaped into the original microchannel dimensions (e), giving the sculpted fluid flow shape. \label{transition_demo}}
\end{figure}

Once the transition matrices for individual pillar transformations have been computed, the net deformation caused by an arbitrary sequence of pillars is easily computed as the matrix product (in sequence) of the corresponding transition matrices. This is schematically shown in Fig.~\ref{transition_demo}. In this example, the inlet flow shape is represented as a vector $\mu_{inlet}$. The fluid is transformed by a sequence of three pillars, which have transition matrices $P_1$, $P_2$, and $P_3$, respectively. Then, the outlet flow shape $\mu_{outlet}$ is given by 
\begin{equation*}
\mu_{outlet}=\mu_{inlet}P_{1}P_{2}P_{3}
\end{equation*}
Thus, the prediction of fluid flow shapes from arbitrary micropillar sequences has been reduced from computationally expensive CFD to simple sparse matrix multiplication \cite{Note1}. The key advantages of the formalism are the speed with which flow shapes can be predicted (see Fig.~\ref{ga_time_res}(a)), and the subsequent simplicity of integrating this with a multitude of optimization frameworks. Furthermore, sparse matrices require far less memory than a dense matrix of similar size, since only the non-zero values are stored (along with their locations in the matrix). In our case, note that each cell is displaced to a single cell (a binary transition). Thus, the framework optimally deploys sparsity with the number of non-zero values in each row equal to 1. This reduces computational overhead. In the next section, we explore the accuracy of this transition matrix representation as a function of discretization.

\subsection{Implementation}

\begin{figure}[t]
\includegraphics[width=\textwidth, keepaspectratio]{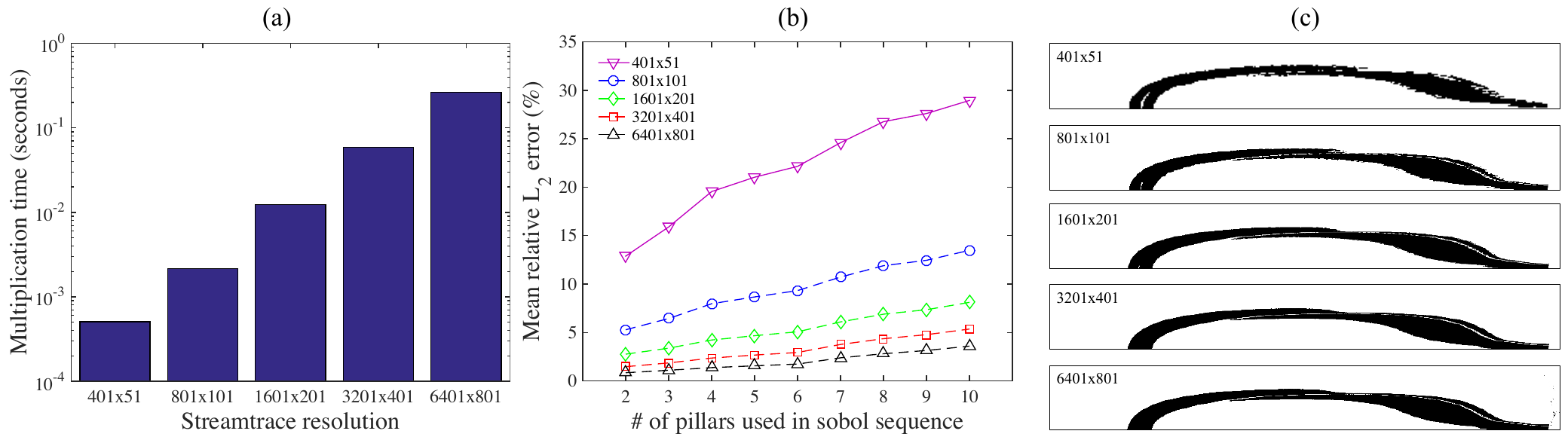}
\caption{(a) Log-scale time for the sparse matrix-matrix product of two random sparse matrices, averaged over 1,000 multiplications, per resolution of transition matrix.  (b) Comparison of truncation error vs number of matrix multiplications, averaged over a 100 sample sobol sequence for each number of possible pillars.  (c) A 10-pillar outlet flow shape for different levels of streamtrace discretization.  Note that these images only show the top half of the microchannel cross-section.  \label{ga_time_res}}
\end{figure}

Consider a transition matrix constructed from a discretization of the microchannel crosssection into $N_Y \times N_Z$ uniform cells. The cells have dimensions $\frac{w}{N_Y}\times\frac{h}{N_Z}$, but are effectively represented as points in the transition matrix.  As such, subtle behavior of streamtraces will be truncated to displacements that align with the discretized representation.  For example, fluid movements too small to leave their immediate cell space will not register in the transition matrix (see cell 1 in Fig.~\ref{truncation_error}(a)). Similarly, movement just large enough to arrive in a new cell will be considered as completely displaced to this cell center (see cell 2 in Fig~\ref{truncation_error}(a). The precision of the transition matrix is therefore limited by the level of discretization of the cross-section into cells. Increasing the number of cells (larger $N_Y, N_Z$) accounts for smaller displacements, thereby leading to a more accurate mapping onto the transition matrix (see Fig.~\ref{truncation_error}(c)). However, larger discretization results in longer computational times. Fig.~\ref{ga_time_res}(a) plots the time needed for a single matrix multiplication of transition matrices constructed with increasing cell discretization. A doubling of discretization (along both $N_Y$ and $N_Z$) increases the computational time by a factor of around 5 in Matlab running on a 2.0 GHz 8-Core Intel E5 2650. The $801\times 101$ and $1601 \times 201$ based discretizations take about 2 milliseconds, and 10 milliseconds, respectively, which result in reasonable run times for the optimization framework. 

\begin{figure}[t]
\includegraphics[width=\textwidth, keepaspectratio]{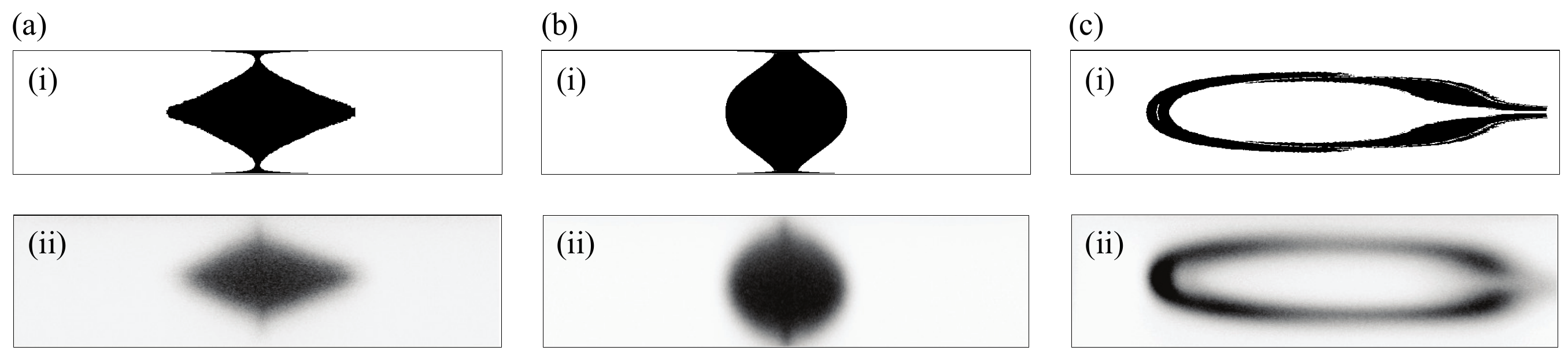}
\caption{Flow deformations \textit{add vertex} (a), \textit{make convex} (b), and \textit{encapsulate} (c) as predicted by matrix multiplication (i), with experimental confocal images (ii).\label{results_validation}}
\end{figure}

We next explore the representation error that these discretizations produce, especially as a function of nested matrix-matrix products. Note that experimental and mass diffusion constraints (please see Appendix A) limit the maximum number of pillars in a sequence to 10. We evaluate the representation error by comparing the predictions of nested matrix products with those of the advection maps for arbitrary pillar sequences of increasing complexity. Error is computed by first converting the transition matrices back into 2D advection maps, and comparing these to the streamtraced advection maps. Thus, for an arbitrary streamtraced advection map $\mathcal{A}$ and matrix-reconstructed map $\mathcal{A_P}$:
\begin{equation}
\text{error}=\frac{\|(\mathcal{A_P}-\mathcal{A})\|_2}{\|\mathcal{A}\|_2}
\end{equation}
We create many sequences of pillars and compute the error. These sequences are constructed using a quasi-random sobol number generator (which is a low discrepancy generator) for maximal coverage of the phase space of pillar combinations. The mean error over 100 realizations is plotted in Fig.~\ref{ga_time_res}(b). From these plots, we chose to utilize a $N_Y = 801, N_Z = 101$ discretization for subsequent analysis, as it provides a good balance between accuracy and computational overhead.  We finally validate this discretization by comparing the prediction of the transition matrix framework with experimental confocal images (see Section IV B,C) of three flow transformations. Fig.~\ref{results_validation} illustrates this promising comparison.


\section{Design problem}

Having established an efficient pathway for quickly evaluating arbitrary sequences of micropillars, we next formulate the problem of identifying micropillar sequences that result in user defined transformations.  

We formulate the design problem as an optimization problem. Specifically, given a target outlet shape $I_{Target}$, we define a cost functional $C (I_{Target}, I_{Test}({\bf s}))$ that returns a scalar value representing how closely a test shape, $I_{Test}({\bf s})$, matches the target shape. $I_{Test}(\bf s)$ is the output transformation of an arbitrary sequence of pillars ${\bf s} \equiv \{s_1,s_2,...,s_k\}$. Each pillar, $s_i$, is chosen from a set of possibilities which are defined in Fig.~\ref{fig:pillar_index_table}. The optimization problem is defined as 
\begin{eqnarray}
\begin{aligned}
\underset{{\bf s} \equiv \{s_1,s_2,...,s_k\}}{\operatorname{argmin}} C(I_{Target}, I_{Test}({\bf s}) ) 
\end{aligned}
\end{eqnarray}
where the pillar sequence ${\bf s}$ that minimizes the fitness function is found. 

\begin{figure}[h]
\includegraphics[width=\textwidth, keepaspectratio]{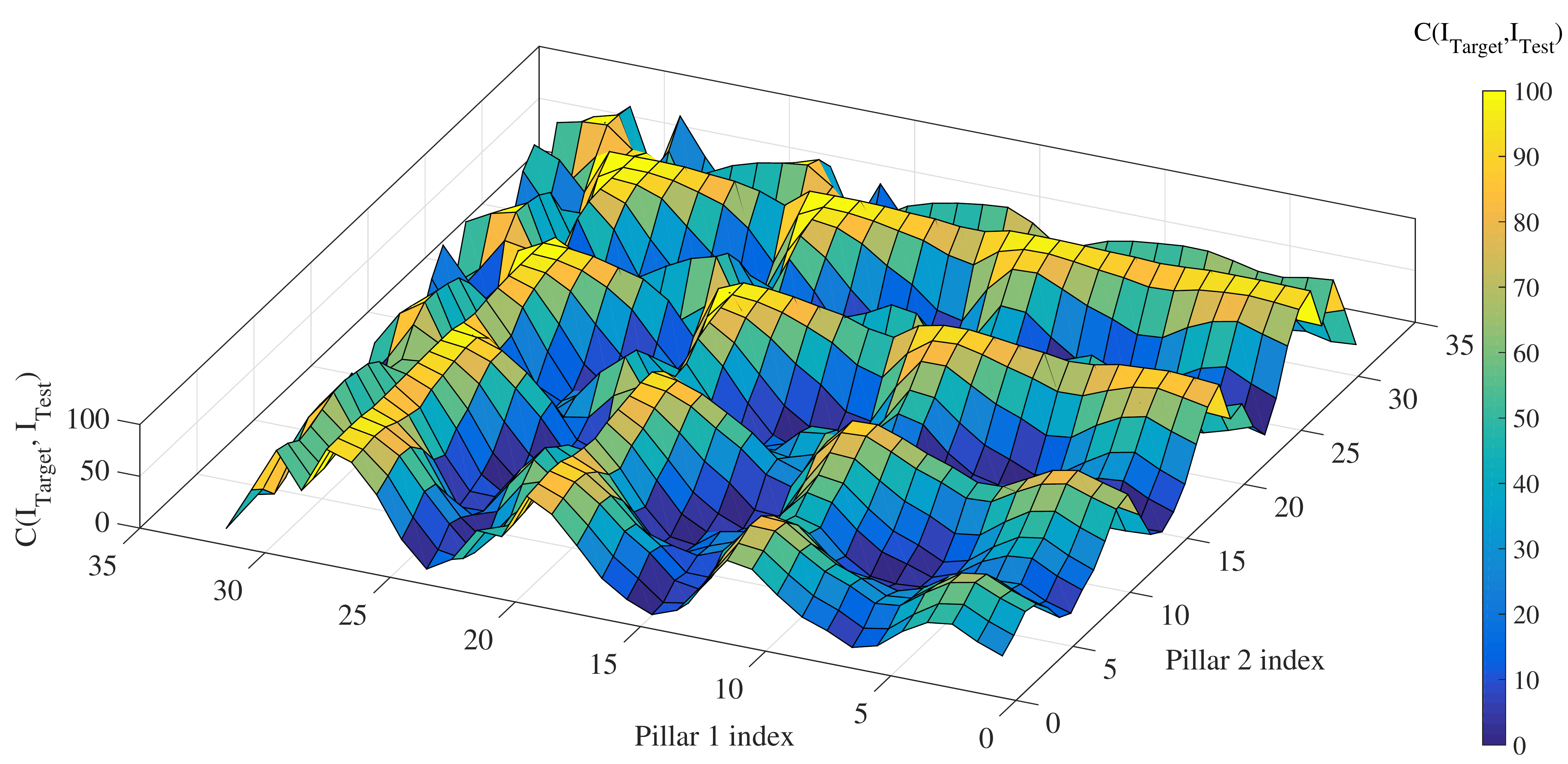}
\caption{This figure shows the design space for a 2-micropillar target fluid flow shape.  The topology will change depending on the target, and quickly become far more complicated in n-dimensions for n-micropillar designs.  Note the multi-modal, corrugated nature of this simple $32 \times 32$ space.\label{design_space}}
\end{figure}

There are several approaches to solve this optimization problem. Here, we choose to utilize a gradient free, evolutionary  optimization strategy. The rationale behind this choice is motivated by the following: 

\begin{enumerate}
\item{The problem formulation as defined here is inherently discrete. This is due to the underlying experimental constraints. Fabrication of arbitrarily sized pillars at continuous locations within practical tolerance and cost is not viable. Effective manufacturing tolerances benefit from a well defined set of micropillar configurations, rather than a continuous space of micropillar diameters and offsets. Simulating a significant subset of this infinite library is also not practical.  Nevertheless, the discrete set from which each pillar is chosen makes the possible search space countably large, hence, precluding an exhaustive search.}
\item{Though it is difficult to illustrate the phase space a 10-pillar sequence offers, a cost function (developed later in this work) evaluated for 2-pillar sequences (with each pillar chosen from 32 possible configurations as shown in Fig.~\ref{fig:pillar_index_table}) is shown in Fig.~\ref{design_space}.  This is a highly corrugated surface, with one global optimum but many nearly identical local minima. This precludes the utilization of gradient based methods, and instead suggests the applicability of stochastic, multistart methods that can explore the phase space efficiently.}
\item{The discrete problem of finding a sequence of matrices such that their product matrix has a desired structure is a variant of the minimum length generator sequence problem \cite{NP_Hard}which has been shown to be a NP-hard problem. This necessitates the utilization of (meta)heuristic optimization methods for efficient solutions.}
\end{enumerate}
These issues naturally suggested using gradient-free metaheuristic evolutionary search algorithms.  We specifically use the genetic algorithm (GA) to locate optimal sequences.  GAs are well suited to multi-modal, highly corrugated solution spaces, especially when the cost function is not easily adapted to gradient-based methods\cite{Mohammadi2004,Giles2000, Holland1992}.

\subsection{Genetic Algorithm}

The GA has been used with substantial success in fluid mechanics with applications ranging from efficient macro- to nano-fluid heat exchanger design\cite{Foli2006, Wang2011,Yang2014}, bluff body flow control\cite{Milano2002}, indoor airflow geometry\cite{Xue2013}, and aircraft design\cite{Ahuja2014}.  Though there are many different flavors and modifications of the GA, the basic algorithm is generally as follows:

\begin{enumerate}
\item{Initialize a population of randomly generated points in the search space. This is the first generation of points. The points are encoded (usually as a binary bitstring) and are called ``chromosomes''.} 
\item{Evaluate each chromosome of the current generation using a fitness function (cost function) specific to the problem.}
\item{Select chromosomes based on their fitness to create a new generation of individuals through crossover and mutation methods.  Good fitness is rewarded by increased chance of selection, which will propagate genetic material to the next generation. There are three basic methods of creating the next generation:}
	\begin{itemize}
		\item{Crossover combines the genetic material in individuals' chromosomes (design parameters) to create offspring that should retain some of the traits of the parents. A crossover rate determines the proportion of each new generation that should come from crossover.}
		\item{Mutation randomly alters the bits within a chromosome.  This introduces random variation into the population, which can bring candidate solutions out of local optima.}
		\item{A set of top-performing chromosomes in the population are chosen as Elites, whose chromosomes are preserved in the next generation.  This retains the optimal solution throughout the GA.}
	\end{itemize}
\item{Repeat steps 2-3 until some set of termination criteria are met, which typically include the following:}
	\begin{itemize}
		\item{Stall generation limit: If mean or optimal fitness does not improve for a set number of generations, terminate the GA.}
		\item{Generation limit: If the number of generations reaches this value, terminate the GA.}
		\item{Stall time: if the GA runtime meets this value, terminate the GA}
	\end{itemize}

\end{enumerate} 

From the final generation, a single optimal solution is selected by choosing the most-fit individual (chromosome).  For this work, we formulate the GA as a minimization problem, making lower fitness more desirable. Because GAs deploy a population of potential solutions distributed over the design space, they are less prone to getting stuck in shallow local minima. GAs are an inherently stochastic method, so we repeat each optimization multiple times (10 times) to consider statistical significance of results and attempt to reliably explore the phase space. 

There are two primary design choices for implementing the GA: chromosome design and choice of the fitness function.  The chromosome used here simply consists of pillar sequences for a fixed number of pillars, with each chromosome bit having an integer value as shown in Fig~\ref{fig:pillar_index_table}.  So, the first generation is a set of completely random pillar sequences of a fixed length.  During the evaluation stage of the GA, the previously described forward model creates a fluid flow shape for each chromosome, which is then compared to the target image in the fitness function.

The choice of the fitness function critically determines the success of the optimization procedure. Since the primary goal of this exercise is to capture the overall shape of the transformation, the fitness function should be a measure of the topology of the shape. Extensive numerical experiments suggested that the (image) correlation coefficient (defined below) gives substantially better results than several standard pixel based norms, while remaining competitive with respect to computational speed.\cite{Note2}  

The correlation coefficient $r$, is defined as
\begin{equation}
r(I_{Target},I_{test})=\frac{\displaystyle{\sum_{i}^{N_Y}\sum_{j}^{N_Z}(I_{test} - \overline{I_{test}})(I_{Target} -\overline{I_{Target}})}}{\sqrt{(\sum_{i}^{N_Y}\sum_{j}^{N_Z}(I_{test} -\overline{I_{test}})^2)(\sum_{i}^{N_Y}\sum_{j}^{N_Z}(I_{Target} -\overline{I_{Target}})^2)}}
\end{equation}

\begin{figure}[t]
\includegraphics[width=\textwidth, keepaspectratio]{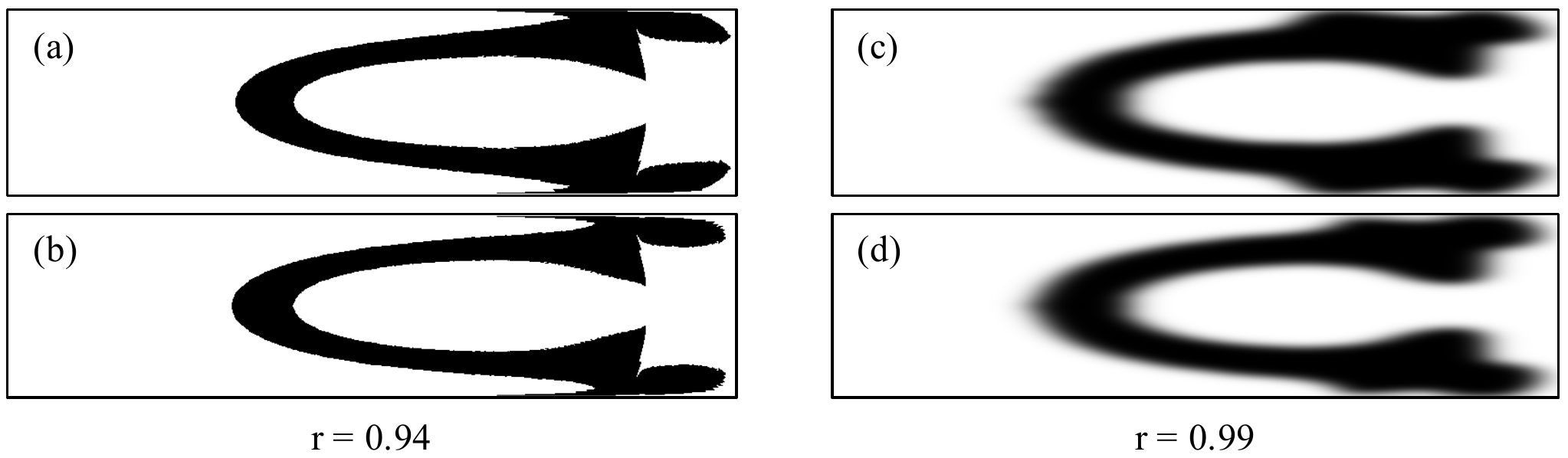}
\caption{Outlet images of 6-pillar (a) and 4-pillar (b) sequences as predicted by transition matrices. Note that despite their overall similarity, minute discrepancies throughout the shape result in a correlation coefficient of $r=0.94$.  When pre-processed by a low-pass filter, as shown in (c) and (d), the correlation coefficient $r=0.99$ relays a more useful comparison of the bulk shapes.\label{fft_lpf_demo}}
\end{figure}

This measure is extensively used to determine how similar two images are.  Identical images would result in $r=1$, while comparing images with precisely complementary pixels would result in $r=-1$.  In the latter case, the result is still desirable, as the overall shape has been achieved using fluid in complementary inlets.  That is, the ``empty'' co-flow inlets have been shaped to align with the target fluid locations in the desired fluid flow shape.  The user would simply need to ``flip'' the inlet flow configuration in order to create their desired microfluidic device. We can treat both optimal values of $r$ equally by squaring its value. Thus, the fitness function $C$ takes the form
\begin{align}
 C & = f(I_{Target}, I_{test}) \\
 C & = 100\times(1-r(I_{Target}, I_{test})^2)
\end{align}
We achieved better convergence by pre-processing $I_{Target}$ and $I_{test}({\bf s})$ before evaluating the fitness function. Specifically, a low-pass filter applied to both $I_{Target}$, and $I_{test}({\bf s})$ emphasized the large scale (topological) features of the shapes while de-emphasizing the fine scale features. This is shown in Fig.~\ref{fft_lpf_demo}. When used with the correlation function, images similar in shape will generally have a better cost functional than without the low-pass filter.  

\section{Experimental Verification of Designed Sequences}

\subsection{Fabrication} All simulations and experiments follow the pillar configurations and flow conditions used in our previous work\cite{Stoecklein2014}, with a microchannel of height $h$, width $w$, microchannel height-to-width aspect ratio $\sfrac{h}{w}=0.25$, and flow conditions $Re = 20$.  The micropillar configurations are shown in Fig~\ref{fig:pillar_index_table}, with indices $1-32$ corresponding to different combinations of normalized pillar diameter $\sfrac{D}{w}=\{0.375,0.5,0.625,0.75\}$, and normalized offset from the center of the channel $\sfrac{y}{w}=\{0.5 (\times2), 0.375, 0.25, 0.125, 0, -0.125, -0.25, -0.375\}$.  The offset location of $\sfrac{y}{w}=0.5 (\times2)$ consists of two half-pillars, located at both sides of the channel.  In fabricated devices, each pillar was spaced approximately 10 pillar diameters apart to ensure deformation saturation.
For verification of the optimized designs, microfluidic chips (200$\mu$m x 50$\mu$m) incorporating the designed pillar sequences were fabricated using soft photolithography. The molds corresponding to the channel design were fabricated from a silicon master spin-coated with KMPR 1050 (MicroChem Corp.) and then patterned by standard photolithography. Polydimethylsiloxane (PDMS) base and curing agent (Sylgard 184 Elastomer Kit, Dow Corning Corporation) were mixed at a ratio of 10 to 1, poured onto the molds in petri dishes, put in a vacuum to remove bubbles, and cured in an oven to replicate the structure of the microchannels. The PDMS devices were peeled from the mold and punched with holes at the inlet and outlet, and bonded with a thin glass slide to enclose the microchannel after activation using air plasma (Plasma Cleaner, Harrick Plasma). The PDMS devices were then filled with Rhodamine B (Sigma-Aldrich), which infused into the PDMS to help visualize the channel walls.

\subsection{Confocal Imaging}
Confocal images of the fluid flow deformation for optimized designs were taken downstream of the fabricated pillars using a Leica inverted SP1 confocal microscope at the California NanoSystems Institute. For each design, three syringes on separate syringe pumps (Harvard Apparatus PHD 2000) were connected to the inlets of the microchannel using PEEK tubing (Upchurch Scientific Product No. 1569). For visualization of the deformed stream, the middle flow stream contained fluorescein isothiocyanate dextran 500kDa (5 $\mu$M, Sigma-Aldrich) while the side streams contained deionized water. The total volume flow rate was 150 $\mu$L/min, with the flow rate of each stream proportional to its cross-sectional area before the first pillar in the design. The confocal images in the cross-sectional plane were taken at least four times of pillar diameter downstream of the last pillar when the flow was fully developed (about 10 minutes after starting to pump). For each measurement, random noise was eliminated by averaging six images to arrive at a final image.

\section{Results}

\begin{figure}[h]
\includegraphics[width=\textwidth, keepaspectratio]{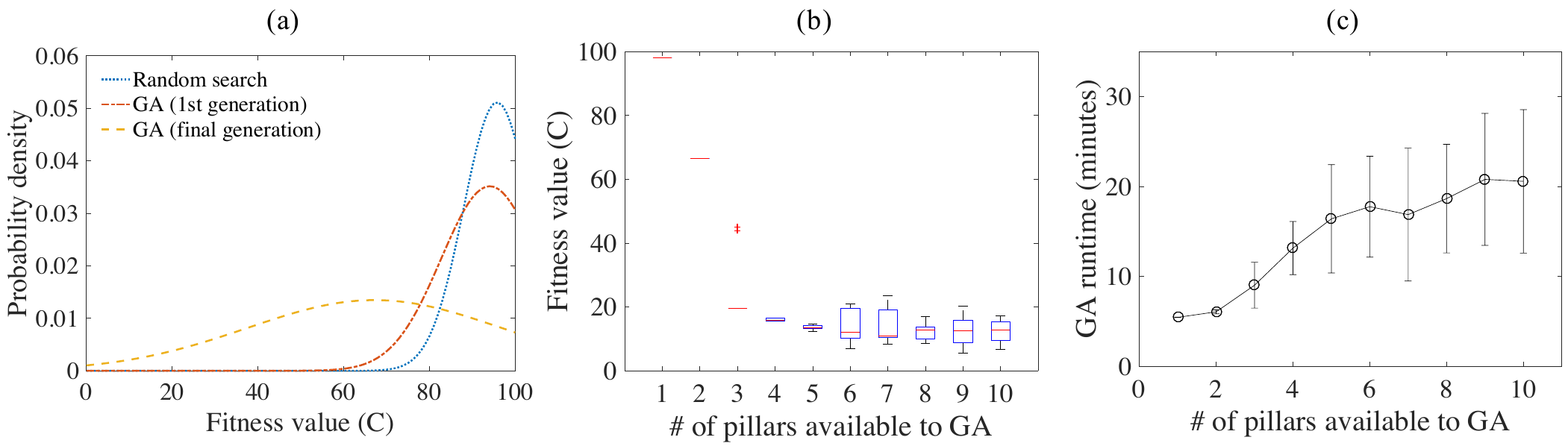}
\caption{(a) Probability density functions for fitness function evaluations for a random search, and the first and last generations of a genetic algorithm.  The random search is based on 10,000 randomly generated sequences with 10 pillars, while the genetic algorithm evolves from an initially random population of 100 10-pillar sequences. Here, the target flow shape was \textit{encapsulate}.  The optimal GA solution had a fitness value $C=6.76$, while the best random solution was $C=21.56$.  (b) Boxplots of optimal fitness values for 10 genetic algorithm trials per number of pillars used in the algorithm chromosomes. Note that the spread of fitness values tends to widen with a larger number of pillars available to the GA, which corresponds to the increasingly complex design space being searched.  (c) Mean runtime for the genetic algorithm based on the low-pass post processed fitness function (see Fig.~\ref{fft_lpf_demo}).  Error bars are the standard deviation for 10 trials per number of pillars available to the genetic algorithm. \label{ga_hist_fitness}}
\end{figure}

\subsection{Improved efficiency of existing designs}

\begin{figure}[t]
\includegraphics[width=\textwidth, keepaspectratio]{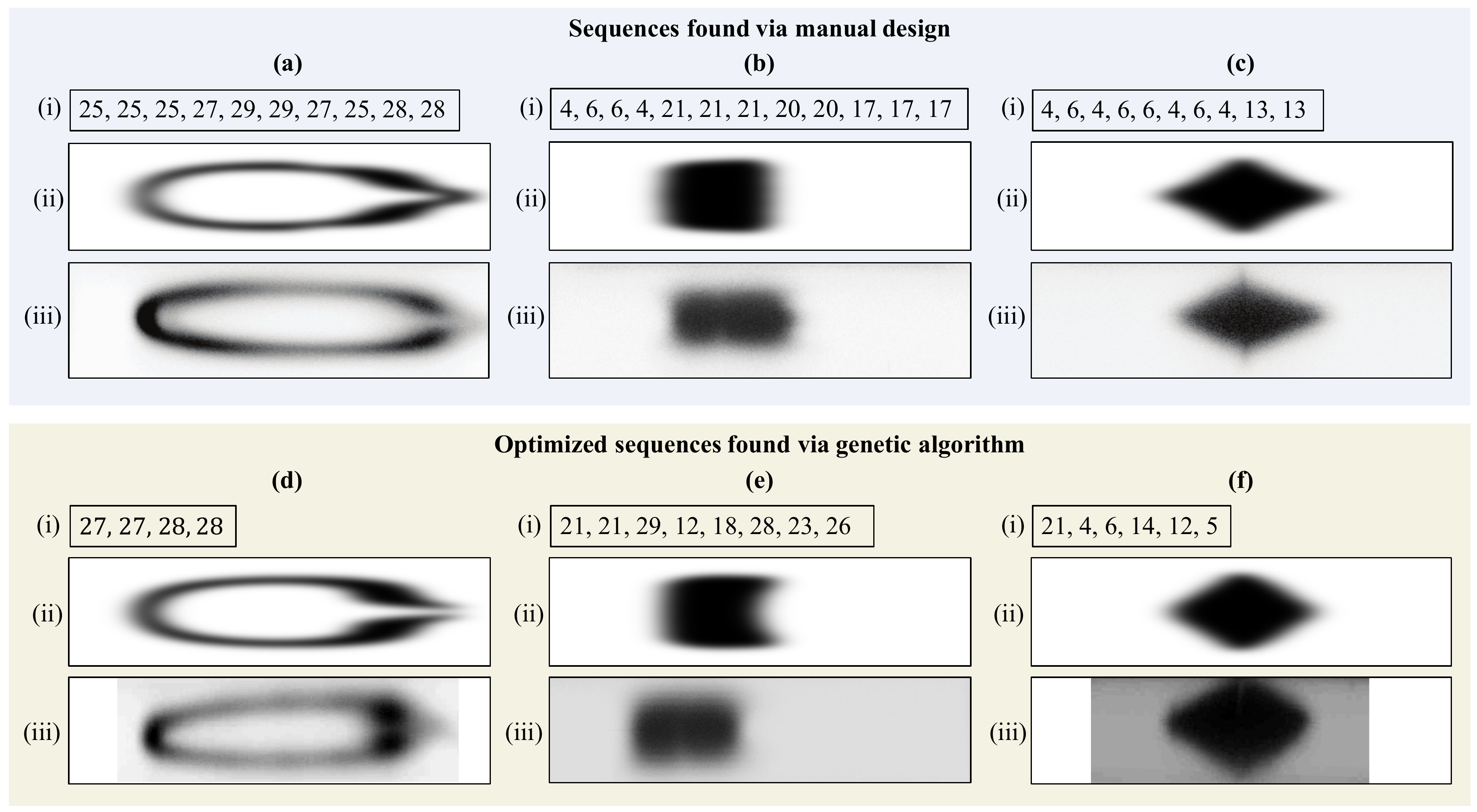}
\caption{Optimization of the 10 pillar \textit{encapsulate} (a), 12 pillar \textit{shift} (b), and 10 pillar \textit{add vertex} (c) transformations  resulted in 4, 8, and 6 pillar optimized sequences (d-f).  Pillar indices for each sequence (i) are found in Fig.~\ref{fig:pillar_index_table}.  Pre-processed transition matrix simulations are seen in (ii), and confocal images of deformed fluid from fabricated devices are shown in (iii). \label{results_enc_shift_av}}
\end{figure}

Three of the more complex, hierarchical designs from our earlier work showing manual designs\cite{Stoecklein2014} were selected to demonstrate the potential to create more efficient pillar sequences for existing transformations. These three transformations are the {\it encapsulate, shift}, and {\it add vertex} transformations (see Fig.~\ref{results_enc_shift_av}(a-c,i-ii)).  All three deform the same inlet flow configuration shown in Fig.~\ref{transition_demo}(a) to entirely different flow shapes. 
The \textit{encapsulate} transformation envelopes a co-flow fluid stream with the primary, central stream.  One example of \textit{encapsulate}'s potential application is seen in a similar shape devised in \cite{Nunes2014}, which formed a new cross-section of a fabricated micropolymer. \textit{Shift} is a powerful manipulation which moves the entire fluid stream across the channel.  Solution transfer\cite{Sollier2015}, enhanced heat transfer, and mixing are examples where \textit{shift} may see application. The {\it add vertex} transformation stands apart due to its sharp vertices at the midsection of the fluid, which may defy expectations of pillar deformation.  \textit{Add vertex} could be the start of interlocking fluid structures, and hints to more complex polygonal fluid flow shapes.

GA optimization found multiple designs for each target transformation, with the shortest sequences and simulated flow shapes shown in Fig.~\ref{results_enc_shift_av}(e-f,i-ii).  See Fig.~\ref{fig:pillar_index_table} for pillar configurations corresponding to the indices in the sequence figures.  The new designs resulted in  33-60\% improvement to sequence length while maintaining the overall desired shape and location in the microchannel.  These designs were fabricated and the pillar transformations were evaluated using confocal images. The experimental results agree well with the predictions as seen in Fig.~\ref{results_enc_shift_av}(d-f,iii).  The improvement by computational optimization also reduced device footprint and pressure by approximately the same percentage in experiments.

All simulations were performed using a Matlab implementation of the sparse matrix operations. We deployed the parallel GA routine available in Matlab on computing clusters.  GA parameters included a population size of 100, crossover rate of 0.8, mutation rate of 0.2, and 5 elites.  Example behavior of the algorithm is shown in Fig.~\ref{ga_hist_fitness}, showing optimum fitness functions and runtime for the optimization of the \textit{encapsulate} flow shape.  We see good support for using GAs by comparing their evolved generation's fitness functions to a computationally equivalent random searches. The GA will evolve 100 initially random pillar sequences for approximately 100 generations (with an upper limit of 200), so we evaluated 10,000 randomly generated pillar sequences.  Fig \ref{ga_hist_fitness}(a) shows fits of normal distributions to the results of these searches, with the GA results having clear improvement over the random search.  The final generation for the GA found an optimum fitness of $C=6.76$, while the random search's best fitness was $C=21.56$.  The GA typically converges on the same optimal shape for smaller design spaces (see Fig \ref{ga_hist_fitness}(b), for \# of pillars = 1-3).  Searches using pillar sequences with a length in excess of 5 pillars find a variety of local optima, which is unsurprising given the larger, more complex search space.  Runtime was as much as 30 minutes, but on average the total run of 10 trials for 10 pillar sequence lengths took about 24 hours on a 2.0 GHz 8-Core Intel E5 2650 CPU.

\subsection{Novel designs}
We next applied the framework to design novel flow transformations. These transformations change the shape of the central fluid into shapes akin to a dumbbell.  The dumbbell shape was motivated by experimental particle fabrication research, where polymer precursors are shaped by a micropillar sequence and subsequently polymerized by UV light.  Use of a shaped mask allows for a single section of flow to be polymerized, rather than the entire stream as in \cite{Paulsen2015}.  Unlike circular or ellipsoid shapes, the dumbbell offers a more complex geometry that could be applied to create particles that align along one axis in a flow\cite{Uspal2013}.  The targets used here differ with respect to the width of the dumbbell, which will show the sensitivity in the choice of a desired shape.

\begin{figure}[t!]
\includegraphics[width=0.8\textwidth, keepaspectratio]{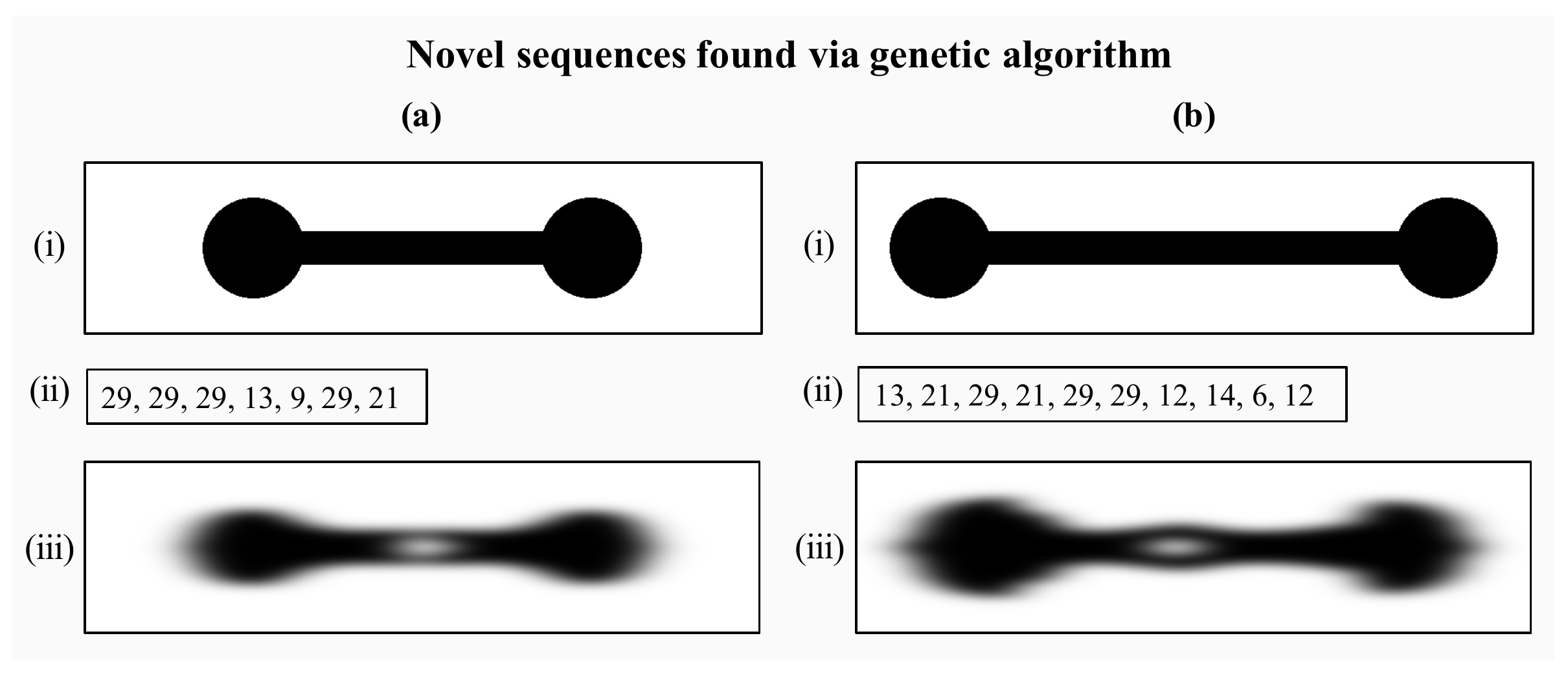}
\caption{Optimization targeting novel ``dumbbell'' shapes (a,i) and (b,i), with genetic algorithm solution sequences (ii) and post-processed flow predictions (iii).  For shape (b,iii), the inlet flow shape spans a width of $\frac{w}{3}$.  Pillar indices can be found in Fig.~\ref{fig:pillar_index_table}. \label{dumbbell_results}}
\end{figure}

The GA was run with several different inlet configurations in order to allow for additional fluid.  Results are in Fig.~\ref{dumbbell_results}, with the wider dumbbell shape requiring an inlet flow shape spanning $\frac{w}{3}$.

\section{Summary}

We have devised and validated an efficient forward model for simulating pillar programming via simple matrix multiplication, and implemented the model into a parallel GA.  The simple nature of the new forward model is valuable for its speed and applicability.  Optimization routines can now incorporate flow shape prediction directly into a computational pipeline.  We have shown how the GA successfully used this new framework by optimizing for known flow shapes, and producing new micropillar sequence designs for novel flow shapes.  Future work on design for pillar programming could pursue faster and more exhaustive searches by tailoring the process to more specialized hardware (e.g., the GPU), and investigating new fitness functions.

Overall, this work completes the cycle moving from a user-defined flow shape of interest to a physical implementation of a channel design that achieves this flow shape. Such an approach opens up the computer-aided design and manufacturing of shaped polymer fibers\cite{Nunes2014} and particles\cite{Paulsen2015} for a range of applications. Additional uses in directing mass and heat transfer, and transferring solutions for automation of biological sample preparation should also benefit. In comparing to experimental and numerical iteration in which the full Navier-Stokes equations are solved for a set of complex channel geometries, our approach achieves orders of magnitudes improvements in time to result, making previously intractable problems that were not even attempted now possible.

\section{Acknowledgements}
This research is supported in part by the National Science Foundation through NSF-1306866, NSF-1307550, and NSF-1149365.

\bibliography{PoF_paper_final.bbl}

\appendix

\section{Mass diffusion limitations}

As the fluid flows downstream, mass diffusion will cause fluid elements to mix according to varying concentrations.  This results in potentially undesirable blurring of the overall shape. This also results in fluid diffusing across streamlines, therefore altering their future trajectories. It is thus important to estimate this diffusive blurring or length scale. We estimate the diffusion length using the relationship between the diffusivity coefficient $D$ and fluid element time of flight, $t$:

\begin{equation*}
\delta=\sqrt{Dt}
\end{equation*}

Our simulations and experiments have a fixed flowrate, with an average velocity $\overline{U}$.  The time of flight can then be determined per number of pillars in a sequence by using the inter-pillar spacing, for which we  use $L = 1.2 mm > 6D_P$ (where $D_P$ is a pillar's diameter) and microchannel width $w=200 \mu m$.  Pillar spacing is critical, as the deformation from one pillar must saturate before before fluid arrives at the subsequent pillar.  Pillars in close proximity to each other ($L < 6D_P$) will encounter ``cross-talk'' in their respective fluid deformation, and the general premise of pillar programming will break down.  Local fluid velocities in the 3D domain will vary depending geometry of the channel, but we can approximate the added time of flight by multiplying the distance travelled by a compensating coefficient $f$:

\begin{equation*}
\delta=\sqrt{\frac{DnfL}{\overline{U}}}
\end{equation*}

Or, we can define the diffusive length using the P\'{e}clet number, $Pe=\frac{w\overline{U}}{D}$:

\begin{equation*}
\delta=\sqrt{\frac{wnfL}{Pe}}
\end{equation*}

\end{document}